\documentclass[aps,prl,showpacs]{revtex4}
\usepackage{graphicx}
\usepackage{amsfonts,amssymb,amsmath}
\usepackage{bm}

\begin{document}

\title{Solution of a Hamiltonian of quantum dots with
Rashba spin-orbit coupling: quasi-exact solution}
\date{\today}
 
\author{Hayriye T\"{u}t\"{u}nc\"{u}ler}
\email{tutunculer@gantep.edu.tr}
\affiliation{Department of Physics, Faculty of Engineering 
University of Gaziantep,  27310 Gaziantep, Turkey}
\author{Ramazan Ko\c{c}}
\email{koc@gantep.edu.tr}
\affiliation{Department of Physics, Faculty of Engineering 
University of Gaziantep,  27310 Gaziantep, Turkey}
\author{Eser Ol\u{g}ar}
\email{olgar@gantep.edu.tr}
\affiliation{Department of Physics, Faculty of Engineering 
University of Gaziantep,  27310 Gaziantep, Turkey}

\begin{abstract}
We present a method to solve the problem of Rashba spin–orbit coupling
in semiconductor quantum dots, within the context of quasi-exactly solvable
spectral problems. We show that the problem possesses a hidden $osp(2,2)$
superalgebra. We constructed a general matrix whose determinant provide
exact eigenvalues. Analogous mathematical structures between the Rashba and
some of the other spin-boson physical systems are notified.
\end{abstract}

\maketitle

\section{Introduction}

The optical and electrical properties of confined electrons in semiconductor
quantum wells, quantum dots and quantum wires depend on the Rashba spin
orbit coupling\cite{rashba,das,winkler}. The analysis of the behavior of
spins in semiconductors leads to the construction of new \textit{spintronic}
devices\cite{prinz,wolf,dresselhaus,wang}. It is practical to contemplate
semiconductor devices based on electron spins, because spin is not coupled
to electromagnetic noise and hence should have much longer coherence times
than charge\cite{band}. The spin-orbit interaction in a confined geometry
has also interesting consequences for the electron spectrum\cite{gover}.

Spin-orbit interactions can arise in quantum dots by various mechanisms
related to the electron confinement and symmetry breaking. In
semiconductors, spin-orbit coupling from the relativistic effect is caused
by the electric field due to the lack of the inversion symmetry of the
certain alloys. Depending on the particular origin of the electric field,
the spin orbit interaction presents two distinct contributions; these are
Dresselhaus and Rashba terms. Dresselhaus term is arised from the electric
field produced by the bulk inversion asymmetry of the material and Rashba
term is generated due to the structural asymmetry of the heterostructure.\
The Rashba splitting has been observed in many experiments and it
constitutes the basis of the proposed electronic nonostructures.

In literature, most of the studies about the solution of the spin-orbit
effects in quantum dots are carried out by means of perturbation theory or
numerical methods, because the implementation of the algebraic techniques to
solve those problems is not very efficient and most of the other analytical
techniques do not yield simple analytical expressions. The aim of this paper
is to provide exact solution of a quantum dot Hamiltonian including Rashba
and Zeeman term by an algebraic formulation. Analytical solution of the
problem has recently been treated by employing various techniques\cite%
{gover,manuel,tsit,diego,sousa}. We demonstrate that the exact
eigenfunctions and eigenvalues are available for the Hamiltonian of a
quantum dot including Rashba coupling in the framework of quasi-exact
solvability(QES)\cite{turbiner,bender,lopez,shif}. As a related topic the
concept of quasi exactly solvable systems discovered in 1980s has received
much attention in recent years, both from the view point of physical
applications and their mathematical beauty. It turns out that in quantum
mechanics there exist such systems; a part of their spectrum can be computed
using algebraic methods. We also show the corresponding Hamiltonian
possesses a hidden $osp(2,2)$ algebraic structure.

Our formulation of the Hamiltonian associated with the quantum dots leads to
an interesting consequence: Hamiltonians of the $E\otimes \varepsilon $
Jahn-Teller\cite{judd,reik,loorits,pooler,koc}, two mode bosonic
Jaynes-Cummings\cite{jaynes,tur,bo} and quantum dot Hamiltonians with
spin-orbit interaction are constructed for similar spin-boson models, though
not identical, and can be solved with the same mathematical techniques. The
connection between the Hamiltonians plays crucial roles to the solution and
analysis of the Rashba Hamiltonian, because the solutions of the Jahn-Teller
and the Jaynes-Cummings problems have been focus of interest both from the
mathematical and physical point of view, over years and their exact solution
has been treated by various authors.

The paper is organized as follows. In section 2, we briefly review the
construction of the Hamiltonian that include Rashba spin-orbit coupling
term. Section 3 is devoted for algebrization of the corresponding
Hamiltonian. In this section we also discuss the symmetry properties of the
Hamiltonian. In section 4, we present a transformation procedure that is
appropriate to determine the QES of the Hamiltonian. Finally, we conclude
our results in section 4.

\section{The model Hamiltonian}

The origin of the Rashba spin-orbit coupling in quantum dots is due to the
lack of inversion symmetry which causes a local electric field perpendicular
to the plane of heterostructure. The Hamiltonian representing the Rashba
spin orbit coupling for an electron in a quantum dot can be expressed as%
\begin{equation}
H_{R}=\frac{\lambda _{R}}{\hbar }\left( p_{y}\sigma _{x}-p_{x}\sigma
_{y}\right)   \label{e1}
\end{equation}%
where $\lambda _{R}$ represents the strength of the spin orbit coupling, and
it can be adjusted by changing the asymmetry of the quantum well via
external electric field and the matrices $\sigma _{x},$ and $\sigma _{y}$
are Pauli matrices. The Hamiltonian dominated by the bulk inversion symmetry
term, is known as Dresselhaus spin orbit coupling and it is given by%
\begin{equation}
H_{D}=\frac{\lambda _{D}}{\hbar }\left( p_{y}\sigma _{y}-p_{x}\sigma
_{x}\right)   \label{e2}
\end{equation}%
where $\lambda _{D}$ is the Dresselhaus parameter that depends on the
material properties, device design and external electric field. In a
centrosymmetric crystal like materials it becomes zero, because of the non
existence of bulk inversion asymmetry. In the case of both Rashba and
Dresselhaus interactions exist in the system the Hamiltonian can not exactly
be solved. During our treatment we have included Rashba term rather than the
Dresselhaus term, because Rashba term may be dominant, since the typical
ratio of the coupling strengths $\lambda _{R}/\lambda _{D}=6$\cite{tsit}. We
mention here that, the results will only need a trivial modification, when a
solo Dresselhaus term is present because Rashba and Dresselhaus terms
transform into each other under spin rotation. To this end we assume that
the electron is confined in a parabolic potential%
\begin{equation}
V=\frac{1}{2}m^{\ast }\omega _{0}^{2}(x^{2}+y^{2})  \label{e3}
\end{equation}%
here $m^{\ast }$ is the effective mass of the electron and $\omega _{0}$ is
the confining potential frequency. The Hamiltonian describing an electron in
two-dimensional quantum dot takes the form%
\begin{equation}
H=\frac{1}{2m^{\ast }}\left( P_{x}^{2}+P_{y}^{2}\right) +\frac{1}{2}g\mu
B\sigma _{z}+V+H_{R}.  \label{e4}
\end{equation}%
The term $\frac{1}{2}g\mu B\sigma _{z}$ introduces the Zeeman splitting
between the $(+)x-$polarized spin up and $(-)x-$polarized spin down. The
factors $g$ is the gyromagnetic ratio and $\mu $ is the Bohr magneton. The
kinetic momentum $\mathbf{P=p+eA}$ is expressed with canonical momentum $%
\mathbf{p}=-i\hbar (\partial _{x},\partial _{y},0)$ and the vector potential 
$\mathbf{A}$ can be related with the magnetic field $\mathbf{B=\nabla }%
\times \mathbf{A}$. The choice of symmetric gauge vector potential $\mathbf{A%
}=B/2(-y,x,0)$, leads to the following Hamiltonian%
\begin{eqnarray}
H &=&-\frac{\hbar ^{2}}{2m^{\ast }}\left( \frac{\partial ^{2}}{\partial x^{2}%
}+\frac{\partial ^{2}}{\partial y^{2}}\right) +\frac{1}{2}m^{\ast }\omega
^{2}(x^{2}+y^{2})+  \notag \\
&&\frac{1}{2}i\hbar \omega _{c}\left( y\frac{\partial }{\partial x}-x\frac{%
\partial }{\partial y}\right) +\frac{1}{2}g\mu B\sigma _{z}+H_{R}  \label{e5}
\end{eqnarray}%
where $\omega _{c}=eB/m^{\ast }$ stands for the cyclotron frequency of the
electron, $\omega =\sqrt{\omega _{0}^{2}+\left( \frac{\omega _{c}}{2}\right)
^{2}}$ is the effective frequency. From now on we restrict ourselves to the
solution of (\ref{e5}). It will be shown that the Hamiltonian (\ref{e5})
without Dresselhaus term is one of the recently discovered quasi-exactly
solvable operator\cite{turbiner,bender}. It is well known that the
underlying idea behind the quasi exact solvability is the existence of a
hidden algebraic structure. In the following section we obtain an algebraic
expression for the Hamiltonian $H$ and discuss its quasi-exact solvability.

\section{QES of the Hamiltonian}

The general procedure to solve a differential equation quasi-exactly is to
express it in terms of the given Lie algebra having a finite dimensional
invariant subspace and use algebraic operations. One way to relate the
Hamiltonian $H$ with an appropriate Lie algebra is to construct its bosonic
and fermionic representation. For this purpose let us introduce the
following bosonic operators:%
\begin{eqnarray}
a_{1}^{+} &=&\sqrt{\frac{m^{\ast }\omega }{4\hbar }}(x+iy)-\sqrt{\frac{\hbar 
}{4m^{\ast }\omega }}(\partial _{x}+i\partial _{y})  \notag \\
a_{1} &=&\sqrt{\frac{m^{\ast }\omega }{4\hbar }}(x-iy)+\sqrt{\frac{\hbar }{%
4m^{\ast }\omega }}(\partial _{x}-i\partial _{y})  \notag \\
a_{2}^{+} &=&\sqrt{\frac{m^{\ast }\omega }{4\hbar }}(x-iy)-\sqrt{\frac{\hbar 
}{4m^{\ast }\omega }}(\partial _{x}-i\partial _{y})  \label{e6} \\
a_{2} &=&\sqrt{\frac{m^{\ast }\omega }{4\hbar }}(x+iy)+\sqrt{\frac{\hbar }{%
4m^{\ast }\omega }}(\partial _{x}+i\partial _{y}).  \notag
\end{eqnarray}%
They satisfy the usual commutation relations. The Hamiltonian $H$ describing
a two-level fermionic subsystem coupled to two boson modes can be expressed
as:%
\begin{eqnarray}
H &=&\hbar \omega (a_{1}^{+}a_{1}+a_{2}^{+}a_{2}+1)+\frac{\hbar \omega _{c}}{%
2}\left( a_{1}^{+}a_{1}-a_{2}^{+}a_{2}\right) -  \notag \\
&&\sqrt{\frac{m^{\ast }\omega }{\hbar }}\lambda _{R}\left[
(a_{2}^{+}-a_{1})\sigma _{+}+(a_{2}-a_{1}^{+})\sigma _{-}\right] +\frac{1}{2}%
g\mu B\sigma _{0}  \label{e7}
\end{eqnarray}%
The Pauli matrices are given by%
\begin{equation}
\sigma _{+}=\left( 
\begin{array}{cc}
0 & 1 \\ 
0 & 0%
\end{array}%
\right) ;\quad \sigma _{-}=\left( 
\begin{array}{cc}
0 & 0 \\ 
1 & 0%
\end{array}%
\right) ;\quad \sigma _{0}=\left( 
\begin{array}{cc}
-1 & 0 \\ 
0 & 1%
\end{array}%
\right)  \label{e8}
\end{equation}%
Before we turn to explore the solvability of the Hamiltonian $H$ we briefly
discuss its Lie (super)algebraic properties.

The natural step to relate a Hamiltonian with a Lie (super)algebra is to
express the Hamiltonian with the generators of the relevant symmetry group.
One of the major symmetry group candidates for spin one-half particles is
the supergroup $osp(2,2)$ which has four even and four odd generators. Its
even generators can be represented by bosons while odd generators are
represented by combinations of the fermions and bosons\cite{chen1,chen2}.
The Lie superalgebra $osp(2,2)$ can be constructed by extending $su(1,1)$
algebra whose generators are given by%
\begin{equation}
J_{+}=a_{1}^{+}a_{2}^{+},\quad J_{-}=a_{2}a_{1},\quad J_{0}=\frac{1}{2}%
\left( a_{1}^{+}a_{1}+a_{2}^{+}a_{2}+1\right) .  \label{e9}
\end{equation}%
These are the Schwinger representation of $su(1,1)$ algebra and its number
operator is given by,%
\begin{equation}
N=a_{1}^{+}a_{1}-a_{2}^{+}a_{2}  \label{e10}
\end{equation}%
which commutes with the $su(1,1)$ generators. The superalgebra $osp(2,2)$
might be constructed by extending $su(1,1)$ algebra with the fermionic
generators 
\begin{equation}
V_{+}=\sigma _{+}a_{2}^{+},V_{-}=\sigma _{+}a_{1},W_{+}=\sigma
_{-}a_{1}^{+},W_{-}=\sigma _{-}a_{2}.  \label{e11}
\end{equation}%
The total number operator of the system is given by 
\begin{equation}
J=\frac{1}{2}\left( N-\sigma _{0}\right) .  \label{e12}
\end{equation}%
The conserved quantity of a physical system possesses $osp(2,2)$
superalgebra can be written as 
\begin{equation}
K=N-\frac{1}{2}\sigma _{0}  \label{ee1}
\end{equation}%
which commutes with the operators of the $osp(2,2)$ superalgebra. The
generators of the $osp(2,2)$ superalgebra satisfy the following commutation
and anti-commutation relations: 
\begin{eqnarray}
\left[ J_{+},J_{-}\right]  &=&-2J_{0},\quad \quad \left[ J_{0},J_{\pm }%
\right] =\pm J_{\pm },  \notag \\
\left[ J,J_{\pm }\right]  &=&0,\quad \quad \quad \quad \left[ J,J_{0}\right]
=0  \notag \\
\left[ J_{0},V_{\pm }\right]  &=&\pm \frac{1}{2}V_{\pm },\quad \left[
J_{0},W_{\pm }\right] =\pm \frac{1}{2}W_{\pm },\quad   \notag \\
\left[ J_{\pm },V_{\mp }\right]  &=&V_{\pm },\quad \quad \quad \left[ J_{\pm
},W_{\mp }\right] =W_{\pm },\quad \quad   \notag \\
\left[ J,W_{\pm }\right]  &=&-\frac{1}{2}W_{\pm },\quad \quad \left[
J,V_{\pm }\right] =\frac{1}{2}V_{\pm }  \label{e13} \\
\left[ J_{\pm },V_{\pm }\right]  &=&0,\quad \quad \quad \quad \left[ J_{\pm
},W_{\pm }\right] =0  \notag \\
\left\{ V_{\pm },W_{\pm }\right\}  &=&J_{\pm },\quad \quad \left\{ V_{\pm
},W_{\mp }\right\} =\pm J_{0}-J  \notag \\
\left\{ V_{\pm },V_{\pm }\right\}  &=&\left\{ V_{\pm },V_{\mp }\right\} =0 
\notag \\
\left\{ W_{\pm },W_{\pm }\right\}  &=&\left\{ W_{\pm },W_{\mp }\right\} =0. 
\notag
\end{eqnarray}

The Hamiltonian of a physical system, with an underlying $osp(2,2)$
symmetry, has been expressed in terms of the operators of the corresponding
algebra. In general, the Hamiltonian is exactly solved and the spectrum of
the physical system can be calculated in a closed form when the Hamiltonian
of the system can be written in terms of number operator and/or diagonal
operator $J_{0}$, or it can be diagonalized within the representation $[J]$.
The abstract boson and/or fermion algebra can be associated with the exactly
solvable Schr\"{o}dinger equations by using the differential operator
realizations of boson operators. This connection opens the way to an
algebraic treatment of a large class of potentials of practical interest\cite%
{gursey}. The combinations of the operators of $osp(2,2)$ superalgebra have
direct physical meaning, being related to the quantum spin systems.

The Hamiltonian (\ref{e7}) can be expressed in terms of operators of $%
osp(2,2)$ algebra

\begin{equation}
H=2\hbar \omega J_{0}+\frac{\hbar \omega _{c}}{2}N-\kappa \left[
V_{+}-V_{-}+W_{-}-W_{+}\right] +g\mu B\left( N-K\right) .  \label{e14}
\end{equation}%
where $\kappa =\sqrt{\frac{m^{\ast }\omega }{\hbar }}\lambda _{R}$.
Consequently we have shown that the Rashba Hamiltonian possesses the $%
osp(2,2)$ symmetry. The Hamiltonian (\ref{e14}) has $(2N+1$) linearly
independent eigenfunctions, and it is quasi-exactly solvable.

Now we turn our attention to the solution of the Hamiltonian (\ref{e7}). A
simple connection between the Hilbert space and the Bargmann-Fock space can
be obtained by transforming the differential realizations of the creation
and annihilation operators (\ref{e6}). This can be done by introducing the
following similarity transformation operators 
\begin{equation}
\Lambda =\exp \left[ -\frac{\pi }{4}\left(
a_{1}^{+}a_{2}+a_{2}^{+}a_{1}\right) \right] ;\quad \Gamma =\exp \left[ 
\frac{\pi }{8}\left( a_{1}^{2}+a_{1}^{+2}-a_{2}^{2}-a_{2}^{+2}\right) \right]
.  \label{e16}
\end{equation}%
The similarity transformation and change of the variable $y\rightarrow iy$
give the following realizations%
\begin{eqnarray}
b_{1} &=&\Gamma \Lambda a_{1}\Lambda ^{-1}\Gamma ^{-1}=\sqrt{\frac{\hbar }{%
m^{\ast }\omega }}\frac{\partial }{\partial x},\quad b_{1}^{+}=\Gamma
\Lambda a_{1}^{+}\Lambda ^{-1}\Gamma ^{-1}=\sqrt{\frac{m^{\ast }\omega }{%
\hbar }}x  \notag \\
b_{2} &=&\Gamma \Lambda a_{2}\Lambda ^{-1}\Gamma ^{-1}=\sqrt{\frac{\hbar }{%
m^{\ast }\omega }}\frac{\partial }{\partial y},\quad b_{2}^{+}=\Gamma
\Lambda a_{2}^{+}\Lambda ^{-1}\Gamma ^{-1}=\sqrt{\frac{m^{\ast }\omega }{%
\hbar }}y.  \label{e17}
\end{eqnarray}%
It is obvious that the operator $K$ commutes with the whole Hamiltonian and
the eigenvalue problem%
\begin{equation}
K\left| n_{1,}n_{2}\right\rangle =\left( k+\frac{1}{2}\right) \left|
n_{1,}n_{2}\right\rangle ,  \label{e15}
\end{equation}%
in the Bargmann-Fock space leads to the following solution:%
\begin{equation}
\psi (x,y)=x^{k}\phi \left( xy\right) \left| \uparrow \right\rangle
+x^{k+1}\phi \left( xy\right) \left| \downarrow \right\rangle .  \label{e18}
\end{equation}%
where $\left| \uparrow \right\rangle $ stands for up state and $\left|
\downarrow \right\rangle $ stands for down state. The eigenfunction of the
Hamiltonian can be obtained from the relation%
\begin{equation}
\left| n_{1},n_{2}\right\rangle =\Lambda ^{-1}\Gamma ^{-1}\psi (x,y).
\label{e19}
\end{equation}%
Substitution of (\ref{e18}) into the Hamiltonian (\ref{e7}) leads to the
following set of one variable differential equations 
\begin{subequations}
\begin{align}
\hbar \omega \left[ 2z\frac{d}{dz}+k+1+\frac{k\omega _{c}}{2\omega }-\frac{%
\mu gB}{2\hbar \omega }-\frac{E}{\hbar \omega }\right] \phi _{1}\left(
z\right) +\lambda _{R}\left[ k+1-\frac{m^{\ast }\omega z}{\hbar }+z\frac{d}{%
dz}\right] \phi _{2}\left( z\right) & =0  \label{e20a} \\
\hbar \omega \left[ 2z\frac{d}{dz}+k+2+\frac{(k+1)\omega _{c}}{2\omega }+%
\frac{\mu gB}{2\hbar \omega }-\frac{E}{\hbar \omega }\right] \phi _{2}\left(
z\right) +\lambda _{R}\left[ \frac{m^{\ast }\omega }{\hbar }-\frac{d}{dz}%
\right] \phi _{1}\left( z\right) & =0  \label{e20b}
\end{align}%
where $z=xy$ and $E$ is the eigenvalues of the Hamiltonian $H$ and $\phi
_{1}\left( z\right) $ and $\phi _{2}\left( z\right) $ correspond up and down
eigenstates of the Hamiltonian $H$, respectively. Following the Reik's
analysis\cite{reik} which was constructed to obtain the solution of the $%
E\times \varepsilon $ Jahn-Teller Hamiltonian, one can obtain the isolated
exact solution of the differential equations. Here we present solution of
the problem in the framework of the quasi-exactly solvable problem.

\section{Solution of the Hamiltonian}

In the previous section we have formulated the Rashba Hamiltonian based on
the two boson operators and we have discussed its transformation to the one
variable differential equation. In this section we present a transformation
procedure which leads to the quasi-exact solution of the Hamiltonian (\ref%
{e7}). The Hamiltonian has been characterized by two boson operator. It is
possible to transform the Hamiltonian that can be characterized by one boson
operator. The connection between two and single boson Hamiltonian is given
by a similarity transformation induced by the metric\cite{gursey,koc2}: 
\end{subequations}
\begin{equation}
S=(a_{2}^{+})^{a_{1}^{+}a_{1}+\sigma _{+}\sigma _{-}}\equiv \left( \sigma
_{-}\sigma _{+}+\sigma _{+}\sigma _{-}a_{2}^{+}\right)
(a_{2}^{+})^{a_{1}^{+}a_{1}}  \label{e21}
\end{equation}%
The operator $(a_{2}^{+})^{a_{1}^{+}a_{1}}$ acts on the state $\left|
n_{1},n_{2}\right\rangle $ as follows,

\begin{equation}
(a_{2}^{+})^{a_{1}^{+}a_{1}}\left| n_{1},n_{2}\right\rangle
=(a_{2}^{+})^{n_{1}}\left| n_{1},n_{2}\right\rangle =\sqrt{\frac{n_{2}!}{%
(n_{2}+n_{1})!}}\left| n_{1},n_{2}+n_{1}\right\rangle  \label{e22}
\end{equation}%
then we can easily obtain the action of the operator $S$ on the two
component state%
\begin{eqnarray}
&&S\left( \left| n_{1},n_{2}\right\rangle \left| \uparrow \right\rangle
+\left| n_{1},n_{2}\right\rangle \left| \downarrow \right\rangle \right) = 
\notag \\
&&\sqrt{\frac{(n_{2}+n_{1})!}{n_{2}!}}\left( \sqrt{(n_{2}+n_{1}+1)}\left|
n_{1},n_{2}+n_{1}+1\right\rangle \left| \uparrow \right\rangle +\left|
n_{1},n_{2}+n_{1}\right\rangle \left| \downarrow \right\rangle \right) .
\label{e23}
\end{eqnarray}%
Since $a_{1}$ and $a_{2}$ commute, the transformation of $a_{1}$ and $%
a_{1}^{+}$ under $S$ can be obtained by writing $a_{2}^{+}=e^{b}$, with $%
[a_{1},b]=[a_{1}^{+},b]=0$,

\begin{equation}
S^{-1}a_{1}S=a_{1}a_{2}^{+};\quad S^{-1}a_{1}^{+}S=a_{1}^{+}(a_{2}^{+})^{-1}
\label{e24}
\end{equation}%
and transformation of $a_{2}$ and $a_{2}^{+}$ as follows

\begin{equation}
S^{-1}a_{2}S=a_{2}+(a_{1}^{+}a_{1}+\sigma _{+}\sigma
_{-})(a_{2}^{+})^{-1};\quad S^{-1}a_{2}^{+}S=a_{2}^{+}.  \label{e25}
\end{equation}%
Similarly the transformation of the Pauli matrices are%
\begin{equation}
S^{-1}\sigma _{+}S=\sigma _{+}(a_{2}^{+})^{-1};\quad S^{-1}\sigma
_{-}S=\sigma _{-}a_{2}^{+};\quad S^{-1}\sigma _{0}S=\sigma _{0}  \label{e26}
\end{equation}%
The Hamiltonian $H$ can be transformed%
\begin{eqnarray}
\widetilde{H} &=&S^{-1}HS=\hbar \omega
(2a_{1}^{+}a_{1}+a_{2}^{+}a_{2}+\sigma _{+}\sigma _{-}+1)-\frac{\hbar \omega
_{c}}{2}\left( a_{2}^{+}a_{2}+\sigma _{+}\sigma _{-}\right) -  \notag \\
&&\kappa \left[ (1-a_{1})\sigma _{+}+(a_{2}a_{2}^{+}+a_{1}^{+}a_{1}+\sigma
_{+}\sigma _{-}-a_{1}^{+})\sigma _{-}\right] +\frac{1}{2}g\mu B\sigma _{0}
\label{e27}
\end{eqnarray}%
It is obvious that $\widetilde{H}$ can be characterized by a fixed number $%
a_{2}^{+}a_{2}=-j-1$. Here $j$ takes integer values. The transformed
Hamiltonian $\widetilde{H}$ includes one boson operator and possessing
infinitely many finite-dimensional invariant subspaces with a basis function:%
\begin{equation}
\phi _{n,n+1}(z)=\left( 
\begin{array}{c}
p_{n}(z) \\ 
q_{n+1}(z)%
\end{array}%
\right) .  \label{e28}
\end{equation}%
where $\phi (z)$ is two component spinor and $p_{n}(z)$ and $q_{n+1}(z)$ are
polynomials of degree $n$ and $n+1$ respectively. Substituting Bargmann-Fock
space realizations of the bosonic operators (\ref{e17}) by changing variable 
$z=\sqrt{\frac{m^{\ast }\omega }{\hbar }}x,$ we obtain the following single
variable differential equation:%
\begin{eqnarray}
\widetilde{H} &=&\hbar \omega \left( 2z\frac{d}{dz}-j+\sigma _{+}\sigma
_{-}\right) +\frac{\hbar \omega _{c}}{2}\left( j+1-\sigma _{+}\sigma
_{-}\right) +g\mu B\frac{\sigma _{0}}{2}-  \notag \\
&&\kappa \left[ \left( 1-\frac{d}{dz}\right) \sigma _{+}+\sigma _{-}(z\frac{d%
}{dz}-j-z)\right] .  \label{e29}
\end{eqnarray}%
The eigenvalue problem can be expressed as%
\begin{equation}
\widetilde{H}\phi (z)=E\phi (z)  \label{e30}
\end{equation}%
The action of the $\widetilde{H}$ on the basis function $\phi (z)$ gives the
following recurrence relation: 
\begin{eqnarray}
\left[ 2n\hbar \omega -E+\epsilon _{j}+\epsilon _{b}\right] p_{n}(E)-\kappa %
\left[ q_{n+1}(E)-(n+1)q_{n}(E)\right]  &=&0  \notag \\
\left[ 2n\hbar \omega -E+\epsilon _{j}-\epsilon _{b}\right]
q_{n+1}(E)+\kappa \left[ p_{n+1}(E)+(j-n)p_{n}(E)\right]  &=&0  \label{e31}
\end{eqnarray}%
where $p_{n}(E)$ and $q_{n}(E)$ are the coefficients of the polynomials $%
p_{n}(z)$ and $q_{n}(z)$, respectively. The justified parameters are given
by 
\begin{eqnarray}
\epsilon _{j} &=&\hbar \omega \left( \frac{1}{2}-j+\frac{(3+2j)\omega _{c}}{%
4\omega }\right)   \notag \\
\epsilon _{b} &=&\left( \frac{\hbar \omega _{c}}{4}-\frac{\mu gB}{2}+\frac{%
\hbar \omega }{2}\right)   \label{e32}
\end{eqnarray}%
It is necessary that the determinant of these sets of \ be equal to zero
giving the compatibility conditions that establish the locations of the
exact eigenvalues on the energy baseline. The recurrence relation implies
that the wavefunction is itself the generating function of the energy
polynomials. If $E_{k}$ is the roots of the (\ref{e31}), the eigenfunction
truncated at $n=j$ and it is the exact eigenvalues of the Hamiltonian $%
\widetilde{H}.$ These recurrence relations can be written in the matrix form:%
\begin{equation}
\left[ 
\begin{array}{cccccc}
E_{+} & -\kappa  & 0 & 0 & 0 & \cdot  \\ 
j\kappa  & E_{-} & \kappa  & 0 & 0 & \cdot  \\ 
0 & 2\kappa  & E_{+}+2\hbar \omega  & -\kappa  & 0 & \cdot  \\ 
0 & 0 & (j-1)\kappa  & E_{-}+2\hbar \omega  & \kappa  & \cdot  \\ 
0 & 0 & 0 & 3\kappa  & E_{+}+3\hbar \omega  & \cdot  \\ 
\cdot  & \cdot  & \cdot  & \cdot  & \cdot  & \ddots 
\end{array}%
\right] \left[ 
\begin{array}{c}
p_{0} \\ 
q_{1} \\ 
p_{1} \\ 
q_{2} \\ 
p_{2} \\ 
\vdots 
\end{array}%
\right] =0  \label{e33}
\end{equation}%
where $E_{\pm }=\epsilon _{j}\pm \epsilon _{b}-E.$ It is obvious that the
determinant of the matrix forms polynomials in $E$ and the first few of them
are given by%
\begin{align}
D_{0}(E)& =E_{+}E_{-}  \notag \\
D_{1}(E)& =\left( E_{-}+2\hbar \omega \right) \left[ E_{+}E_{-}(E_{+}+2\hbar
\omega )-\kappa ^{2}(E_{+}-2\hbar \omega )\right]   \notag \\
D_{2}(E)& =\left( E_{-}+4\hbar \omega \right) [E_{+}E_{-}(E_{+}+2\hbar
\omega )(E_{-}+2\hbar \omega )(E_{+}+4\hbar \omega )  \label{e34} \\
& +\kappa ^{2}(2\hbar \omega E_{+}(E_{-}+4\hbar \omega )+162\hbar ^{2}\omega
^{2}(E_{-}+2\hbar \omega )-2E_{+}^{2}E_{-})  \notag \\
& +2\kappa ^{4}(E_{+}-2\hbar \omega )]  \notag
\end{align}%
for the values $j=0,1$ and $2$, respectively. The eigenvalues $E$ of the
Hamiltonian (\ref{e7}) can be determined by finding the roots of the
polynomials(\ref{e34}).

\section{Conclusion}

We presented a QES to the problem of an electron in a quantum dot in the
presence of both the magnetic field and spin-orbit coupling. Our formulation
gives a unified treatment of the solution of the spin-boson physical
systems. It has been shown that the mathematical structures of the quantum
dot Hamiltonian including Rashba spin-orbit interaction and some of the
spin-boson physical systems are identical. We have also shown that the
Hamiltonian possesses $osp(2,2)$ hidden symmetries. The suggested approach
can be modified to solve the quantum dot Hamiltonian including Dresselhaus
interaction.

Furthermore we have presented a transformation procedure that offers several
advantageous, especially if one wishes to describe the eigenvalues of the
Hamiltonian. It is obvious that the algebraic techniques have been used in a
variety of problems to compute their spectrums. We have presented the steps
towards an extension of the algebraic formulation of the quantum dot
Hamiltonians.

The technique given in this article can be extended in several ways. The
Hamiltonian of a quantum dot including position dependent effective mass may
be formulated and solved within the procedure given here. We hope that our
method leads to interesting results on the spin-orbit effects in quantum
dots in future research.


\bigskip

\begin{thebibliography}{99}
\bibitem{rashba} Bychkov Y A and Rashba E I, 1984 \textit{J. Phys.C} \textbf{%
17} 6039

\bibitem{das} Datta S and Das B 1990 \textit{Appl. Phys. Lett.} \textbf{56}
665

\bibitem{winkler} Winkler R 2000 \textit{Phys. Rev. B} \textbf{62} 4254

\bibitem{prinz} Prinz G A 1998 \textit{Science} \textbf{282} 1660

\bibitem{wolf} Wolf S A, Awschalom D D, Buhrman R A, Daughtan J M, vov Moln%
\={a}r S, Roukes M L, Chtchelkanova A Y and Treger D M 2001 \textit{Science} 
\textbf{294} 1488

\bibitem{dresselhaus} Dresselhaus G 1955 \textit{Phys. Rev.} \textbf{100} 580

\bibitem{wang} Wang X F, Vasilopoulos P, Peeters F M 2002 \textit{Appl.
Phys. Lett.} \textbf{80} 1400

\bibitem{band} Bandyopadhyay S 2000 \textit{Phys. Rev. B} \textbf{61} 13813

\bibitem{gover} Governale M 2002 \textit{Phys. Rev. Lett.} \textbf{89} 206802

\bibitem{manuel} Val\`{\i}n-Rodr\`{\i}guez M, Puente A and Serra L 2003 
\textit{Eur. Phys. J.} \textbf{B34}, 359

\bibitem{tsit} Tsitsishvili E, Lozano G and Gogolin A O 2003
http://arxiv.org/cond-mat/0310024

\bibitem{diego} Frustaglia D and Richter K 2003
http://arxiv.org/cond-mat/0309228

\bibitem{sousa} de Sousa R and Sarma D S 2003 \textit{Phys. Rev. B} \textbf{%
68} 155330

\bibitem{turbiner} Turbiner A V and Ushveridze A G 1987 \textit{Phys. Lett. A%
} \textbf{126} 181.

\bibitem{bender} Bender C M and Dunne G V,1996 \textit{J. Math. Phys.} 
\textbf{37,} 6

\bibitem{lopez} Gonzales-Lopez A, Kamran N and Olver P J 1993 \textit{%
Comm.Math. Phys.} \textbf{153} 117.

\bibitem{shif} Shifman M A 1989 \textit{Int. J. Mod. Phys.} A\textbf{4}
2897.\qquad

\bibitem{judd} Judd B R 1979 \textit{J. Phys. C} \textbf{12} 1685.

\bibitem{reik} Reik H G, St\"{u}lze M E and Doucha M 1987 \textit{J. Phys. A}
\textbf{20}, 6327.

\bibitem{loorits} Loorits V 1983 \textit{J. Phys. C} \textbf{16}, L711.

\bibitem{pooler} Pooler D R 1978 \textit{J. Phys. A: Math. Gen.} \textbf{11}
1045

\bibitem{koc} Ko\c{c} R, T\"{u}t\"{u}nc\"{u}ler H, Koca M and K\"{o}rc\"{u}k
E 2003 \textit{Prog. Theor. Phys.} \textbf{110} 399

\bibitem{jaynes} Jaynes E T and Cummings F W 1963 \textit{Proc. IEEE} 
\textbf{51} 89

\bibitem{tur} Tur E A 2000 \textit{Opt. Spectrosc.} \textbf{89} 574

\bibitem{bo} Jing-Bo Z U and Xu-Bo Z O U 2001 \textit{Chin. Phys. Lett.} 
\textbf{18} 51

\bibitem{chen1} Chen Yong-Qing 2000 \textit{J. Phys. A: Math. Gen.} \textbf{%
33} 8071; 2001 \textit{Int. J. Theor. Phys.} \textbf{40} 1113; 2000 \textit{%
Int. J. Theor. Phys.} \textbf{39} 2523

\bibitem{chen2} Chen Yong-Qing, Xiao-Hui Liu and Xing-Chang Song 1994 
\textit{Commun. Theor. Phys.} \textbf{22} 123

\bibitem{gursey} Alhassid Y, G\"{u}rsey F and Iachello F 1983 \textit{Ann.
Phys.(N. Y.)} \textbf{148} 346

\bibitem{koc2} T\"{u}t\"{u}nc\"{u}ler H and Ko\c{c} R 2004 \textit{Pramana
J. Phys.} \textbf{62} 993
\end{thebibliography}
\end{document}